%% file: main.tex
\documentclass[journal]{IEEEtran}

\usepackage{graphicx}
\usepackage{amsmath,amssymb}
\usepackage[utf8]{inputenc}
\usepackage{cite}
\usepackage{url}
\usepackage{xcolor}
\usepackage{etoolbox}
\newcommand{\rev}[1]{#1}

\begin{document}
\bstctlcite{SCEPnoURL}

\title{\rev{A Deep Neural Network based Level-1 Trigger for the Magnetic Monopole Search in the SCEP Experiment}}

\input{authorlist_neural_trigger}
\maketitle


\begin{abstract}
The Search for Cosmic Exotic Particles (SCEP) experiment proposes a multilayer \rev{magnetic monopole (MM)} detection array with a target exposure of $O(10^4)~\mathrm{m^2\cdot year}$. Continuously recording the raw waveforms at this scale would generate approximately $O(10)~\mathrm{EB}$ of data per year, making full waveform storage impractical. A Level-1 (\rev{L-1}) trigger is therefore required to reduce the stored-data volume while retaining rare \rev{MM} signals. \rev{MM} waveforms vary with particle velocity and trajectory, so a single filter kernel cannot match all possible signals. A mismatch between the kernel and the signal waveform can reduce the detection probability. This work therefore develops a \rev{compact} template-bank construction method for \rev{MM} detection. Each incoming waveform is filtered with all kernels in the bank, and the best-matched kernel is identified from their responses. A multilayer perceptron (MLP) trigger is further developed to combine all template responses and improve signal detection. With a 31-sample window and a stored-data fraction of $10^{-3}$, the MLP increases the mean net acceptance $P_{\mathrm{net}}$ from \rev{$6.72\%$ for the conventional template-bank trigger} to $8.79\%$, a relative improvement of \rev{$30.8\%$}. The network is then implemented on a Xilinx AC701 board equipped with a field-programmable gate array (FPGA) to process a 20-bit input waveform sampled at $1~\mathrm{MHz}$ using $200~\mathrm{MHz}$ trigger logic. At the target stored-data fraction, the fixed-point implementation reduces $P_{\mathrm{net}}$ by only $3/(3.2\times10^6)$ relative to the floating-point reference. On-board tests reproduce the bit-accurate reference trigger outputs and sustain continuous $1~\mathrm{MHz}$ processing without dropped input samples, verifying the functional correctness and real-time operation of the FPGA-based \rev{L-1} trigger.
\end{abstract}


\section{Introduction}
\label{sec:introduction}

\IEEEPARstart{M}{agnetic} monopoles (MMs), first proposed by Dirac in 1931~\cite{dirac1931quantised}, are hypothetical particles carrying isolated magnetic charge. Their existence would explain electric-charge quantization through the Dirac quantization condition and is also predicted in various grand unified theories~\cite{georgi1974unity,preskill1984magnetic}\rev{,~\cite{thooft1974monopoles}}. Searches for MMs have used several detection strategies. Induction-based searches with superconducting coils look for the persistent current change induced by an MM passage~\cite{eberhard1975improvements,cabrera1982first}. Other searches exploit the energy loss of MMs in detector media. The Monopole, Astrophysics and Cosmic Ray Observatory (MACRO) used scintillation counters, limited streamer tubes, and nuclear track detectors~\cite{macro2002final}. The IceCube Neutrino Observatory has searched for relativistic MMs through their Cherenkov-light signatures in ice~\cite{abbasi2013search}. These techniques have placed strong limits on the \rev{MM} flux~\cite{patrizii2015status,patrizii2019searches}, but their operating conditions or detector-scale requirements make further exposure increases challenging.

Room-temperature induction coils offer a scalable approach to large-area \rev{MM} searches. Unlike superconducting systems, they do not require cryogenic cooling, which facilitates deployment in large-area, multilayer detector arrays. The Search for Cosmic Exotic Particles (SCEP) experiment proposes a coincidence-detection scheme combining room-temperature induction coils with plastic scintillators. Its proposed multilayer array targets an exposure of $O(10^4)~\mathrm{m^2\cdot year}$. Under the detector-performance assumptions of the proposal, this exposure would allow sensitivity comparable to existing \rev{MM}-flux limits over a broad velocity range~\cite{PhysRevD.111.023020}.

For a detector system of this scale, recording all waveform data continuously would overburden the bandwidth and storage capacity of the data-acquisition system. For an array designed to accumulate the target exposure over ten years, the annual raw-data volume is estimated to be $O(10)~\mathrm{EB}$ for the SCEP design. Direct storage of all waveforms is therefore impractical, and a Level-1 (\rev{L-1}) trigger is required to suppress noise-induced events while retaining potential signals for subsequent analysis.

Because MM events are extremely rare and a true event cannot be recovered once the front-end trigger rejects it, the \rev{L-1} trigger must maintain high signal acceptance. In room-temperature induction-coil detection, coil thermal noise results in a low signal-to-noise ratio (SNR) for \rev{MM}-induced signals. For low-velocity \rev{MMs}, a processed-signal SNR of approximately $4.5$ still leaves the signal vulnerable to noise fluctuations~\cite{PhysRevD.111.023020}.
Maintaining high signal acceptance therefore requires a low trigger threshold, which produces a large number of noise-induced false triggers. The stored data are expected to be dominated by noise-induced false triggers rather than genuine signals. The data reduction is quantified by the stored-data fraction, defined as the number of saved waveform samples divided by the total number of raw input samples. A target stored-data fraction of $10^{-3}$ ($\approx10~\mathrm{PB/year}$ for the SCEP design) is adopted as a manageable storage load. In addition, the \rev{L-1} trigger should be implemented in front-end hardware to make trigger decisions in real time and thereby reduce the data-transfer load~\cite{akerib2016luxtrigger}\rev{,~\cite{duarte2018fast}}.

The \rev{L-1}-trigger design for \rev{MM} detection can draw on the approach used in gravitational-wave searches. Induction-based \rev{MM} searches in SCEP and compact-binary gravitational-wave searches with the Laser Interferometer Gravitational-Wave Observatory (LIGO)\rev{~\cite{aasi2015advanced,abbott2016observation}} share a common challenge: both seek weak transient signals in noise-dominated data streams, while the signal parameters are unknown before detection. For a known signal waveform and noise covariance, optimal filtering \rev{(OF)} maximizes the output SNR, whereas a mismatch between the filter kernel and the signal causes an SNR loss\rev{~\cite{turin1960matched}}. A standard way to address unknown signal parameters is to filter each data window with a bank of candidate templates and use the template with the largest response magnitude~\cite{cokelaer2007template,allen2012findchirp,babak2013binary}. Early template-bank constructions set the template spacing through a relative loss in SNR~\cite{PhysRevD.53.6749,PhysRevD.60.022002}. Direct comparisons have shown that adding a more complex template family does not necessarily improve detection performance at a fixed false-alarm rate~\cite{vandenbroeck2009spinning}. Adding templates can increase the largest noise response. At a fixed \rev{false-positive probability (FPR)}, this can raise the trigger threshold and reduce the signal acceptance\rev{~\cite{croce2004correlator}}. Template-bank construction must therefore satisfy the detection-probability coverage requirement while using as few templates as possible. In addition, retaining only the largest template response discards information in the remaining responses that may help distinguish signals from noise.

To address these limitations, this paper constructs a finite template bank using the absolute loss in detection probability as the coverage criterion. This criterion avoids unnecessary refinement of weakly detectable waveform regions. A multilayer perceptron (MLP) trigger is then developed to combine all template responses at each sampling time, improving signal acceptance at the same stored-data fraction. The remainder of this paper is organized as follows. Section~\ref{sec:mismatch_analysis} describes the template bank, the MLP trigger, and the performance evaluation. Section~\ref{sec:fpga_implementation} describes the streaming \rev{field-programmable gate array (FPGA)} architecture for the selected network and its validation procedure. Section~\ref{sec:conclusion} concludes the paper.


\section{\rev{L-1} Trigger Algorithm}
\label{sec:mismatch_analysis}

\subsection{Signal and Noise}
\label{sec:signal_model}

In the SCEP induction-coil scheme, the \rev{MM} signal waveform is set by the particle velocity and trajectory. Each trajectory can be parameterized in a coordinate system fixed to the induction coil. The coil plane is taken as the $xy$-plane, and the $z$ axis is perpendicular to this plane. The physical parameters are denoted by
\begin{equation}
\gamma \equiv (v,\rho_0,\theta,\phi,q),
\qquad q\in\{-1,+1\},
\label{eq:gamma_def}
\end{equation}
where $v$ is the \rev{MM} velocity, $\rho_0$ is the transverse distance from the coil center when the \rev{MM} reaches the coil plane, and $\theta$ and $\phi$ are the polar and azimuthal angles of the directed incident trajectory. The parameter $q$ denotes the magnetic-charge polarity, with $q=+1$ and $q=-1$ corresponding to magnetic charges of $+g_D$ and $-g_D$, respectively. Here, $g_D$ is the unit Dirac magnetic charge\rev{~\cite{dirac1931quantised,patrizii2015status}}.

No well-established theoretical prior is available for the \rev{MM} velocity. Accordingly, $v$ is sampled uniformly on a logarithmic scale over $[10^{-5},10^{-1}]\,c$, where $c$ is the speed of light. This choice prevents the sampled trajectories from being concentrated at high velocities. The dimensionless velocity $\beta\equiv v/c$ is used below. The incoming \rev{MMs} are assumed to be isotropic. The transverse offset $\rho_0$ and the incident angles $(\theta,\phi)$ are not sampled independently; they are determined by the geometrical relation between each isotropic trajectory and the induction coil. The magnetic-charge polarity $q$ is sampled with equal probabilities for $+1$ and $-1$. The signal parameter space of the trajectories parameterized in Eq.~\eqref{eq:gamma_def} is denoted by $\Gamma$.

\begin{figure}[htp]
    \centering
    \includegraphics[width=0.98\columnwidth]{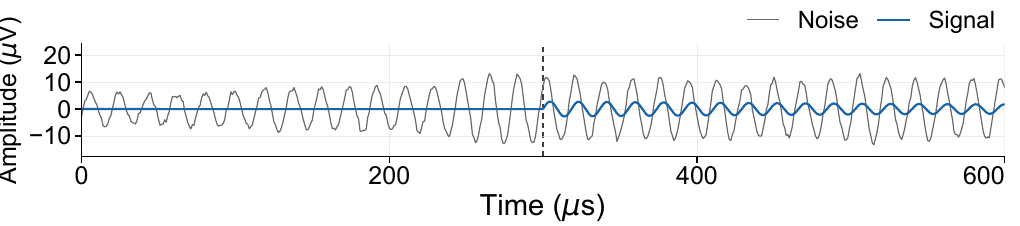}
    \caption{Circuit-shaped \rev{MM} signal and noise. The blue curve shows the circuit-shaped \rev{MM} signal, the gray curve shows the total noise (coil thermal noise plus readout-electronics contributions such as amplifier noise), and the dashed line marks the signal injection time in the analysis window.}
    \label{fig:circuit_response}
\end{figure}

The \rev{MM} crossing induces an electromotive force in the coil. The coil is read out by a low-noise amplifier with high input impedance connected in parallel across its terminals, and the resulting circuit transfer function shapes the induced waveform. The signal and noise models use the same coil design and readout-circuit parameters as those specified in the SCEP proposal~\cite{PhysRevD.111.023020}. The circuit-shaped signal waveform and the noise are shown in Fig.~\ref{fig:circuit_response}. Under room-temperature operation, the dominant noise contribution is the coil thermal noise, which can overwhelm the \rev{MM} signal. The total noise is modeled as a zero-mean stationary Gaussian process. Non-stationary or non-Gaussian disturbances in a deployed detector are beyond the scope of this study.

\subsection{Optimal Filter and Mismatch}
\label{sec:optimal_filter_mismatch}

Detecting \rev{an MM} signal in a noisy waveform can be formulated as a statistical hypothesis-testing problem. Each input waveform window contains $L$ samples. Under the null hypothesis, $\mathcal{H}_0$, the input waveform window contains only noise. Under the alternative hypothesis, $\mathcal{H}_1$, the window contains \rev{an MM} signal plus noise. Expressed mathematically,
\begin{equation}
\begin{aligned}
\mathcal{H}_0 &: \ \mathbf{x}=\mathbf{n}, \\
\mathcal{H}_1 &: \ \mathbf{x}=\mathbf{s}_{\gamma}+\mathbf{n},
\end{aligned}
\label{eq:hypotheses}
\end{equation}
where $\mathbf{x}$ is the input waveform window, $\mathbf{s}_{\gamma}$ is the circuit-shaped signal waveform for parameter $\gamma$, and $\mathbf{n}\sim\mathcal{N}(\mathbf{0},C)$ states that the noise vector $\mathbf{n}$ follows a multivariate Gaussian distribution with zero mean and covariance matrix $C$, computed from the noise power spectral density. This is the sampled form of the Gaussian noise model introduced in Section~\ref{sec:signal_model}. For the hypotheses in Eq.~\eqref{eq:hypotheses} with a specified signal waveform, the Neyman--Pearson lemma\rev{~\cite{neyman1933tests}} gives the likelihood-ratio test (LRT) as the optimal decision rule. In Gaussian noise, it reduces to testing the linear statistic~\cite{kay1998detection}
\begin{equation}
\mathbf{s}_{\gamma}^{\top}C^{-1}\mathbf{x}
\mathop{\gtrless}_{\mathcal{H}_0}^{\mathcal{H}_1}
\lambda.
\label{eq:lrt_linear}
\end{equation}
Here $\lambda$ is the decision threshold: when the linear statistic exceeds $\lambda$, the window is assigned to the signal hypothesis $\mathcal{H}_1$; otherwise it is assigned to the noise-only hypothesis $\mathcal{H}_0$. Equation~\eqref{eq:lrt_linear} has the form of an \rev{OF} statistic $\mathbf{h}_{\gamma}^{\top}\mathbf{x}$, where the \rev{OF} kernel takes the form~\cite{somalwar1988transient}
\begin{equation}
\mathbf{h}_{\gamma}
=
a_{\gamma}C^{-1}\mathbf{s}_{\gamma},
\label{eq:of_kernel}
\end{equation}
where $a_{\gamma}$ is a normalization coefficient chosen so that the root-mean-square noise after \rev{OF} is unity.

\begin{figure}[htp]
    \centering
    \includegraphics[width=0.98\columnwidth]{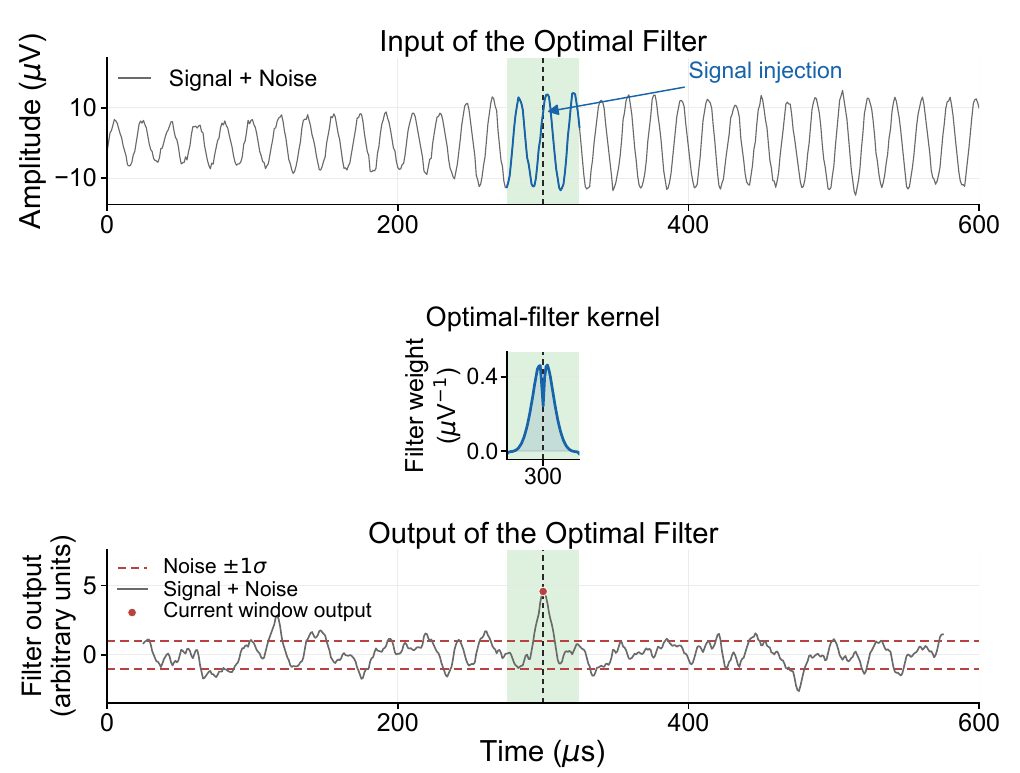}
    \caption{\rev{OF} of a signal-plus-noise waveform. Top: the circuit-shaped input waveform, with the signal injected at the window center marked by the dashed line. Middle: the normalized OF kernel $\mathbf{h}_{\gamma}=a_{\gamma}C^{-1}\mathbf{s}_{\gamma}$ aligned with the signal. Bottom: the OF output, where the matched response forms a peak near the injection time. The red dashed lines indicate the $\pm1\sigma$ noise level after normalization, where $\sigma$ is the noise standard deviation.}
    \label{fig:of_kernel_output}
\end{figure}

In a streaming implementation, at each fixed sampling interval a new sample enters the trigger and the oldest sample leaves, forming the updated input window on which the \rev{OF} is evaluated. The signal is centered within the kernel, and the filter response peaks when the signal in the current window aligns with the template.
At this alignment, the output \rev{SNR} is defined as the squared noise-free signal response divided by the mean squared noise response. With the normalization of the kernel in Eq.~\eqref{eq:of_kernel}, this gives
\begin{equation}
    \mathrm{SNR}(\gamma;\gamma)
    =
    \left(\mathbf{h}_{\gamma}^{\top}\mathbf{s}_{\gamma}\right)^2 .
\label{eq:snr_def}
\end{equation}
The input waveform, \rev{OF} kernel, and resulting filter output are shown in Fig.~\ref{fig:of_kernel_output}.

In practical detection, the true signal parameters $\gamma$ are unknown. When the signal $\mathbf{s}_{\gamma}$ is filtered with the kernel $\mathbf{h}_{\eta}$, where $\eta$ denotes the parameters of the reference signal used to construct the kernel, the resulting output SNR is denoted by $\mathrm{SNR}(\gamma;\eta)$. It is largest for the matched case $\eta=\gamma$, where $\mathrm{SNR}(\gamma;\gamma)$ equals the benchmark of Eq.~\eqref{eq:snr_def}.

\begin{figure}[htp]
    \centering
    \includegraphics[width=0.98\columnwidth]{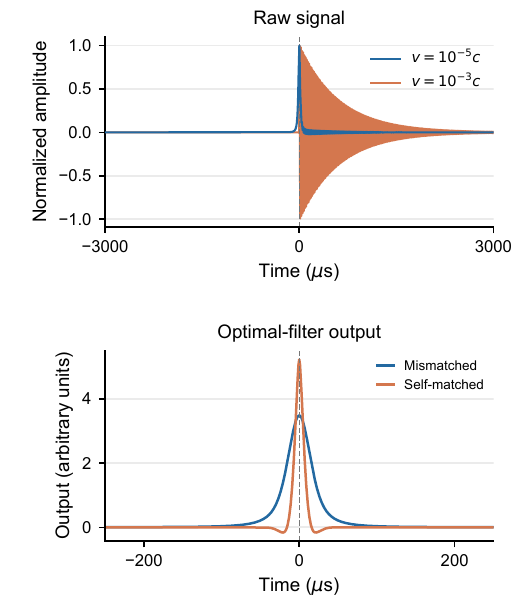}
    \caption{Example of a large template mismatch for two on-axis trajectories with $\rho_0=\theta=\phi=0$. The \rev{upper} panel compares the time-domain waveforms for $v=10^{-5}c$ (blue) and $v=10^{-3}c$ (orange); the signals are normalized to facilitate the comparison of the two waveform shapes. The \rev{lower} panel shows that filtering the $v=10^{-3}c$ signal with the mismatched $v=10^{-5}c$ template (blue) yields a much smaller peak than filtering it with its self-matched template (orange), corresponding to an SNR loss of $55.6\%$.}
    \label{fig:mismatch_example}
\end{figure}

The $\mathrm{SNR}$ loss caused by mismatch can be substantial. Filtering a $v=10^{-3}c$ signal with a $v=10^{-5}c$ template gives an SNR loss of $55.6\%$ relative to the matched case. The corresponding time-domain waveforms and filter-response comparison are shown in Fig.~\ref{fig:mismatch_example}.

\subsection{Template Bank}
\label{sec:mismatch_control}

The preceding example shows that a filter kernel can incur a substantial SNR loss when its waveform does not match the signal. Because the true signal parameters are unknown and the parameter space $\Gamma$ is continuous, a finite template bank $\mathcal{B}$ is constructed to cover the sampled signal distribution within this space. Each input window is filtered with every kernel in $\mathcal{B}$. Templates that differ only in magnetic-charge polarity have opposite kernels, so one kernel is retained for each waveform shape. The template-bank trigger uses the largest response magnitude,
\begin{equation}
\max_{\eta\in\mathcal{B}}
\left|\mathbf{h}_{\eta}^{\top}\mathbf{x}\right|
\mathop{\gtrless}_{\mathcal{H}_0}^{\mathcal{H}_1}
\lambda .
\label{eq:direct_bank_statistic}
\end{equation}

The bank construction has two objectives. First, the templates should cover the sampled signal distribution with a controlled loss in detection probability. Second, the number of templates should be kept small. In addition to increasing the filtering cost, adding templates can increase the probability of a noise-induced trigger under Eq.~\eqref{eq:direct_bank_statistic}.

\begin{figure}[htp]
    \centering
    \includegraphics[width=0.98\columnwidth]{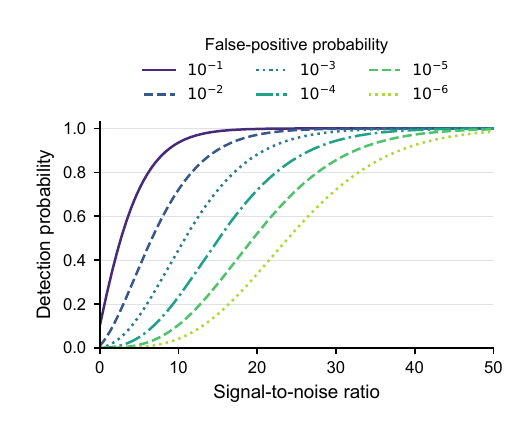}
    \caption{Theoretical single-template detection probability for a normalized Gaussian response as a function of SNR.}
    \label{fig:theoretical_detection_probability}
\end{figure}

Coverage is quantified using the theoretical detection probability for a single normalized template response in Gaussian noise\rev{~\cite{kay1998detection}}. At a specified \rev{FPR}, the detection probability is determined by the signal SNR, as shown in Fig.~\ref{fig:theoretical_detection_probability}. For each signal, its own waveform is used as the ideal matched template and defines the reference detection probability. A signal is considered covered when, at every FPR from $10^{-6}$ to $0.1$, the absolute loss in detection probability obtained with the best-matched template in $\mathcal{B}$ is less than $0.5$ percentage points (pp). \rev{Here, pp denotes the absolute difference between probabilities expressed as percentages; a loss of $0.5$ pp corresponds to a probability difference of $0.005$, not a relative loss of $0.5\%$.} Unlike the relative SNR-loss criterion commonly used in LIGO template-bank construction~\cite{PhysRevD.53.6749,PhysRevD.60.022002}, this absolute criterion relates coverage directly to detectability and prevents signals with self-matched detection probabilities close to the noise-only level from driving further refinement of the bank. To satisfy this coverage criterion while limiting the number of templates, the bank is constructed iteratively using a farthest-first rule~\cite{gonzalez1985clustering}. The construction set is generated using the sampling distribution of Section~\ref{sec:signal_model}. It provides a Monte Carlo representation of the continuous signal parameter space $\Gamma$ under this distribution. At each step, the coverage of the current bank is evaluated over the construction set, and the trajectory with the largest current absolute loss in detection probability is added to the bank. The iteration stops when the loss is below $0.5$ pp for every construction trajectory. By prioritizing the worst-covered trajectory, this rule avoids adding templates for waveforms that are already well represented.

\begin{figure}[htp]
    \centering
    \includegraphics[width=0.98\columnwidth]{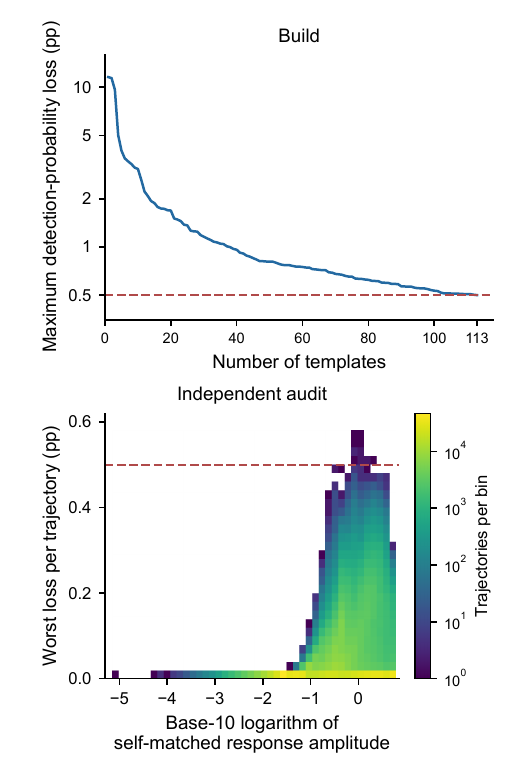}
    \caption{Construction and independent audit of the 113-template bank. Upper: maximum detection-probability loss over the $10^6$ construction trajectories as templates are added. Lower: loss distribution for $10^6$ independent audit trajectories. The dashed lines mark the $0.5$-pp coverage bound. The lower horizontal axis shows the base-10 logarithm of the self-matched response amplitude.}
    \label{fig:coverage_pool}
\end{figure}

The window length $L$ affects the construction result. For $L=31$ samples, the construction stops at 113 templates when all $10^6$ isotropic physical trajectories in the construction set satisfy the $0.5$-pp coverage bound. As shown in the upper panel of Fig.~\ref{fig:coverage_pool}, all construction trajectories satisfy the coverage requirement, with the maximum loss below $0.50$ pp. An independent audit is then performed using another $10^6$ discrete trajectories drawn from the same sampling distribution. The empirical coverage rate exceeds $99.99\%$, with 15 trajectories falling outside the coverage bound; the largest observed loss is $0.57$ pp, as shown in the lower panel of Fig.~\ref{fig:coverage_pool}. These results demonstrate empirical coverage under the stated sampling distribution.

\rev{For comparison, a 601-template bank is constructed for $L=31$ using the stochastic template-placement method employed in LIGO~\cite{harry2009stochastic}, applied to the same MM waveform model. This method controls the relative loss in matched-filter response amplitude by requiring a minimum normalized match of $0.97$, corresponding to a maximum amplitude loss of $3\%$.}

\subsection{Streaming Trigger and Acceptance Evaluation}
\label{sec:stream_evaluation}

\begin{figure}[htp]
    \centering
    \includegraphics[width=0.98\columnwidth]{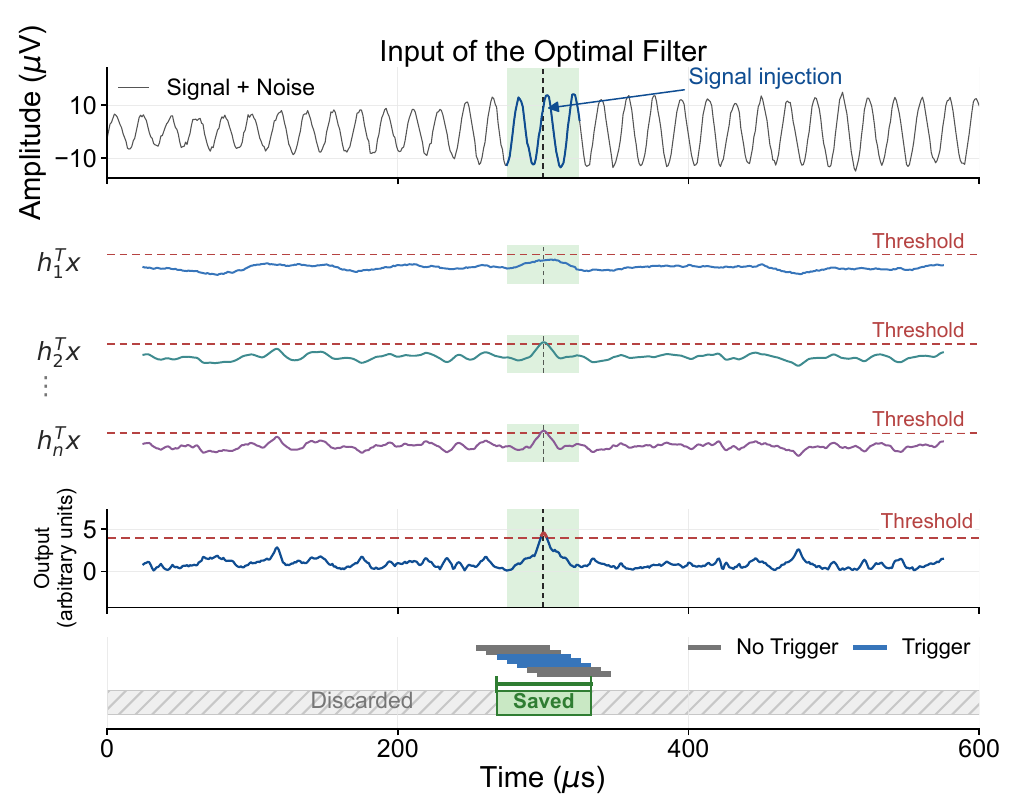}
    \caption{Streaming evaluation of the template-bank trigger. Top: the signal-plus-noise waveform, with the injected signal highlighted around its center time. Middle: representative template responses and the maximum response magnitude, which is compared with the trigger threshold shown by the red dashed line. Bottom: consecutive above-threshold windows are grouped into an event cluster, and their overlapping intervals are merged into the saved waveform segment.}
    \label{fig:stream_cluster}
\end{figure}

After the template bank has been constructed, its practical trigger performance must be evaluated on a continuous waveform stream. Previous gravitational-wave studies have likewise shown that template-space coverage alone does not determine overall detection performance~\cite{croce2004correlator}. At each sampling step, the current $L$-sample window is filtered with every template in the bank, and the largest absolute response in Eq.~\eqref{eq:direct_bank_statistic} is compared with the threshold. When the threshold is exceeded, the corresponding $L$ raw samples are retained. In practice, an event often produces several consecutive above-threshold windows rather than a single isolated trigger. These consecutive windows are grouped into one event cluster. Because the windows are shifted sample by sample, a cluster spanning $N_{\mathrm{trig}}$ consecutive trigger windows retains
\begin{equation}
L_{\mathrm{saved}}
=
L+N_{\mathrm{trig}}-1
\label{eq:l_saved}
\end{equation}
raw samples. If the saved intervals from neighboring clusters overlap, they are merged before storage, so that no raw sample is stored twice. The storage rule in Eq.~\eqref{eq:l_saved}, including merging of overlapping intervals, is illustrated in Fig.~\ref{fig:stream_cluster}.

Under this storage rule, the acceptance of a template bank can be evaluated at a specified stored-data fraction. For each trigger method, the threshold is first calibrated on the negative set $\mathcal{D}_{-}$, taken as a simulated continuous background-only noise stream of $2.01\times10^8$ raw samples. The threshold is chosen so that the stored-data fraction $\varepsilon$ on $\mathcal{D}_{-}$ reaches the target value $\varepsilon_0=10^{-3}$.

\begin{figure}[htp]
    \centering
    \includegraphics[width=0.98\columnwidth]{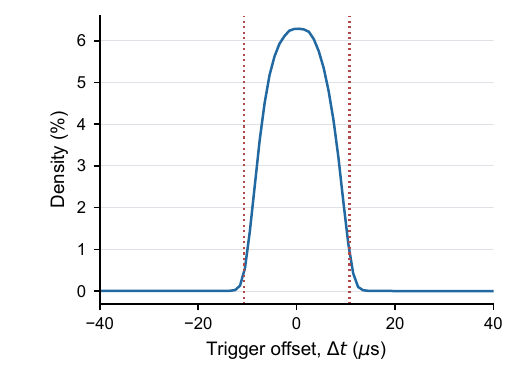}
    \caption{Background-subtracted trigger-offset distribution for the compact template-bank trigger at $\varepsilon_0=10^{-3}$. The dotted lines mark the pooled descriptive value $W_{99}=10.75~\mu\mathrm{s}$.}
    \label{fig:dt_distribution}
\end{figure}

After the threshold is fixed, the corresponding trigger is evaluated on the positive set $\mathcal{D}_{+}$. This set contains $3.2\times10^6$ signal-plus-noise records. Because the \rev{MM} crossing is asynchronous with the sampling clock, each injection is given a random sub-sample timing offset $\delta\in[-0.5,0.5]$ samples, so that the evaluation reflects arbitrary sampling phases. A signal is counted as detected if its event cluster contains at least one trigger sufficiently close to the true crossing time. The allowed timing deviation accounts for the spread introduced mainly by noise fluctuations and the random sub-sample timing offset.
The relevant quantity is the trigger offset
\begin{equation}
\Delta t
=
t_{\mathrm{trig}}-t_0 ,
\label{eq:dt_def}
\end{equation}
where $t_{\mathrm{trig}}$ denotes the center time of an above-threshold trigger window and $t_0$ is the true crossing time. The distribution of $\Delta t$ defined in Eq.~\eqref{eq:dt_def} is obtained from an independent timing-calibration set generated from the same signal distribution as $\mathcal{D}_{+}$. All above-threshold positions are included, and noise-only background is subtracted statistically. For each trigger method and target stored-data fraction, $W_{99}$ is calibrated separately in each velocity bin using the frozen threshold. The symmetric interval $[-W_{99},W_{99}]$ contains $99\%$ of the corresponding background-subtracted distribution. Figure~\ref{fig:dt_distribution} shows the $\Delta t$ distribution of the template-bank trigger at a stored-data fraction of $10^{-3}$. A signal-plus-noise record is classified as correctly triggered if at least one above-threshold position satisfies $|\Delta t|\le W_{99}$; otherwise, it is classified as missed. The net acceptance $P_{\mathrm{net}}$ is the fraction of correctly triggered signal-plus-noise records minus the fraction of paired noise-only records satisfying the same timing criterion.

\begin{figure}[htp]
    \centering
    \includegraphics[width=0.98\columnwidth]{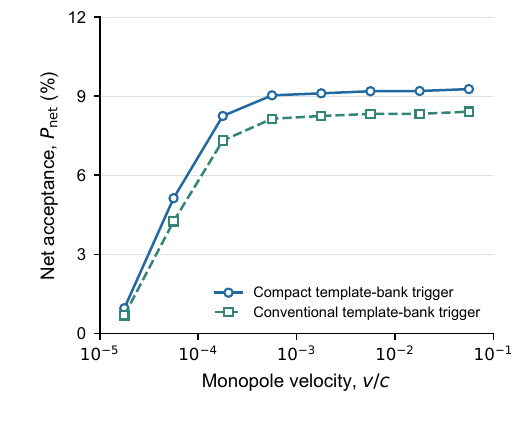}
    \caption{Velocity-dependent net acceptance $P_{\mathrm{net}}$ of the compact \rev{template-bank trigger (113 templates) and the conventional template-bank trigger (601 templates)} at a stored-data fraction of $10^{-3}$. \rev{Both methods use separately calibrated thresholds and timing windows.} The evaluation uses \rev{the same} $3.2\times10^6$ signal-plus-noise records.}
    \label{fig:template_bank_velocity_acceptance}
\end{figure}

Using the calibrated threshold and the timing criterion above, Fig.~\ref{fig:template_bank_velocity_acceptance} shows the velocity-dependent net acceptance $P_{\mathrm{net}}$ of the compact \rev{template-bank trigger and the conventional template-bank trigger} at a stored-data fraction of $10^{-3}$. \rev{For the compact template-bank trigger,} $P_{\mathrm{net}}$ is lowest in the slowest velocity bin and increases from $0.96\%$ to approximately $8.25\%$--$9.27\%$ over the intermediate and higher velocity bins. \rev{The compact template-bank trigger gives higher $P_{\mathrm{net}}$ in every velocity bin. Its mean $P_{\mathrm{net}}$ is $7.52\%$, compared with $6.72\%$ for the conventional template-bank trigger, corresponding to an $11.9\%$ relative improvement while using fewer templates.}

\subsection{MLP Trigger}
\label{sec:neural_trigger}

The compact template-bank trigger uses a practical approximation to the fixed-parameter LRT and therefore does not retain its Neyman--Pearson optimality. Although the template bank provides high empirical coverage under the stated sampling distribution, coverage alone does not guarantee the highest acceptance at a fixed stored-data fraction. Moreover, the compact template-bank trigger retains only the largest response magnitude in Eq.~\eqref{eq:direct_bank_statistic}, discarding the information carried by the remaining template responses. To address these limitations, the MLP trigger is trained to approximate a likelihood-ratio decision rule based on the vector of absolute responses from all templates, thereby learning how to weight and combine all template outputs rather than relying only on their maximum\rev{~\cite{cranmer2015likelihood}}.

\begin{figure}[htp]
    \centering
    \includegraphics[width=0.98\columnwidth]{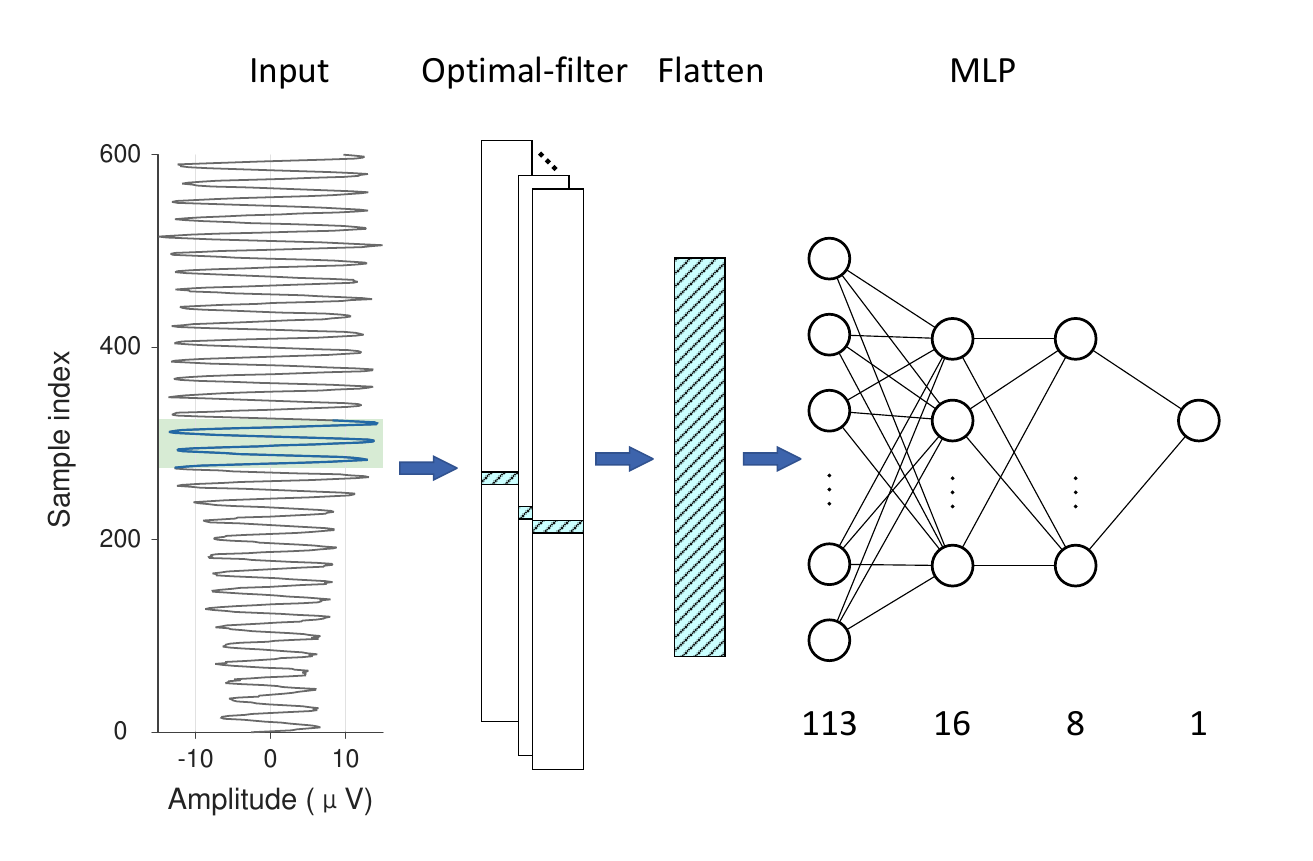}
    \caption{Architecture of the MLP trigger. The absolute responses of all 113 templates form the input vector, which is processed by \rev{FC} layers with 16 and 8 hidden units and one output logit. The hidden layers use ReLU activations.}
    \label{fig:mlp1_architecture}
\end{figure}

At each sampling time, the absolute responses of all filters in the template bank form the input vector to the MLP trigger. The network maps this vector through two fully connected \rev{(FC)} hidden layers with 16 and 8 units, respectively, followed by one output logit. Both hidden layers use the rectified linear unit (ReLU) activation\rev{~\cite{glorot2011rectifier}}, as illustrated in Fig.~\ref{fig:mlp1_architecture}.

The training set contains $2\times10^6$ positive and $4\times10^6$ negative waveform windows. A signal-injected window is labeled positive when its center lies within $3~\mu\mathrm{s}$ of the true crossing time. Pure-noise windows are labeled negative, as are signal-injected windows whose centers are more than $10~\mu\mathrm{s}$ from the true crossing time; signal-injected windows with offsets between $3$ and $10~\mu\mathrm{s}$ are excluded. An independent validation set with the same signal distribution and labeling rules is used to monitor overfitting and select the checkpoint. The networks are trained with ordinary binary cross-entropy.

\begin{figure}[htp]
    \centering
    \includegraphics[width=0.98\columnwidth]{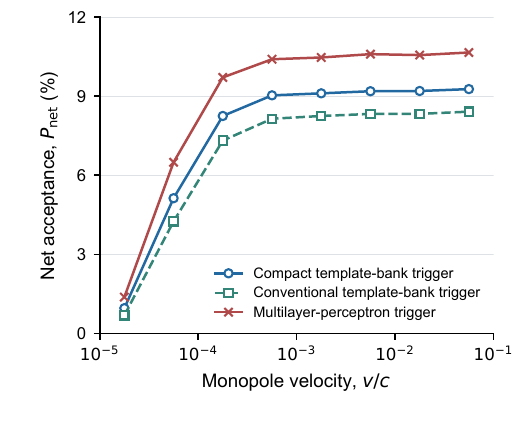}
    \caption{Velocity-dependent net acceptance $P_{\mathrm{net}}$ of the MLP trigger\rev{,} compact \rev{template-bank trigger (113 templates), and conventional template-bank trigger (601 templates)} for $L=31$ at a stored-data fraction of $10^{-3}$. \rev{All three} methods are evaluated using the same $3.2\times10^6$ signal-plus-noise records.}
    \label{fig:mlp_velocity_acceptance}
\end{figure}

Each frozen network is evaluated using the same streaming procedure as in Section~\ref{sec:stream_evaluation}. Figure~\ref{fig:mlp_velocity_acceptance} compares the MLP trigger \rev{with both template-bank triggers} for $L=31$. The MLP trigger provides higher net acceptance $P_{\mathrm{net}}$ in every velocity bin, with increases of $0.42$--$1.46$ pp \rev{over the compact template-bank trigger}. Over the complete evaluation set, $P_{\mathrm{net}}$ increases from $7.52\%$ to $8.79\%$, corresponding to a relative improvement of $16.8\%$.

\begin{figure}[htp]
    \centering
    \includegraphics[width=0.98\columnwidth]{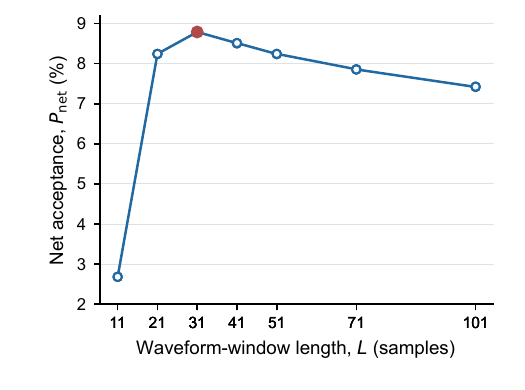}
    \caption{Mean net acceptance $P_{\mathrm{net}}$ of the MLP trigger in the length-selection study as a function of the waveform-window length at a stored-data fraction of $10^{-3}$. All lengths use the same $3.2\times10^6$ signal-plus-noise records. The highlighted marker identifies the selected configuration, $L=31$.}
    \label{fig:mlp_length_selection}
\end{figure}

The waveform-window length creates a tradeoff in stream-level triggering. An above-threshold window causes its corresponding $L$-sample interval to be retained, so increasing $L$ increases the amount of data saved per trigger. In contrast, an excessively short window captures less signal energy and makes signal--noise discrimination more difficult. An intermediate value of $L$ may therefore provide the highest acceptance at the fixed stored-data fraction. The candidate lengths are $L\in\{11,21,31,41,51,71,101\}$ samples. The frozen networks for all lengths are evaluated on the same $3.2\times10^6$ signal-plus-noise records, and the length giving the largest mean net acceptance $P_{\mathrm{net}}$ is selected.
Figure~\ref{fig:mlp_length_selection} summarizes the length scan. The $L=31$ configuration gives the largest mean net acceptance $P_{\mathrm{net}}$ in the length-selection study and is selected for the subsequent FPGA implementation.

\section{FPGA Implementation of the Neural-Network Trigger}
\label{sec:fpga_implementation}

\subsection{Overall Architecture}
\label{sec:fpga_architecture}

To process the continuous waveform stream in real time, the MLP trigger described in Section~\ref{sec:neural_trigger} is implemented on an FPGA. The target input is a 20-bit waveform sampled at $1~\mathrm{MHz}$, while the trigger logic operates at $200~\mathrm{MHz}$ on a Xilinx AC701 board equipped with an XC7A200T FPGA~\cite{xilinxUG952,xilinxDS180}.

\begin{figure}[htp]
    \centering
    \includegraphics[width=0.98\columnwidth]{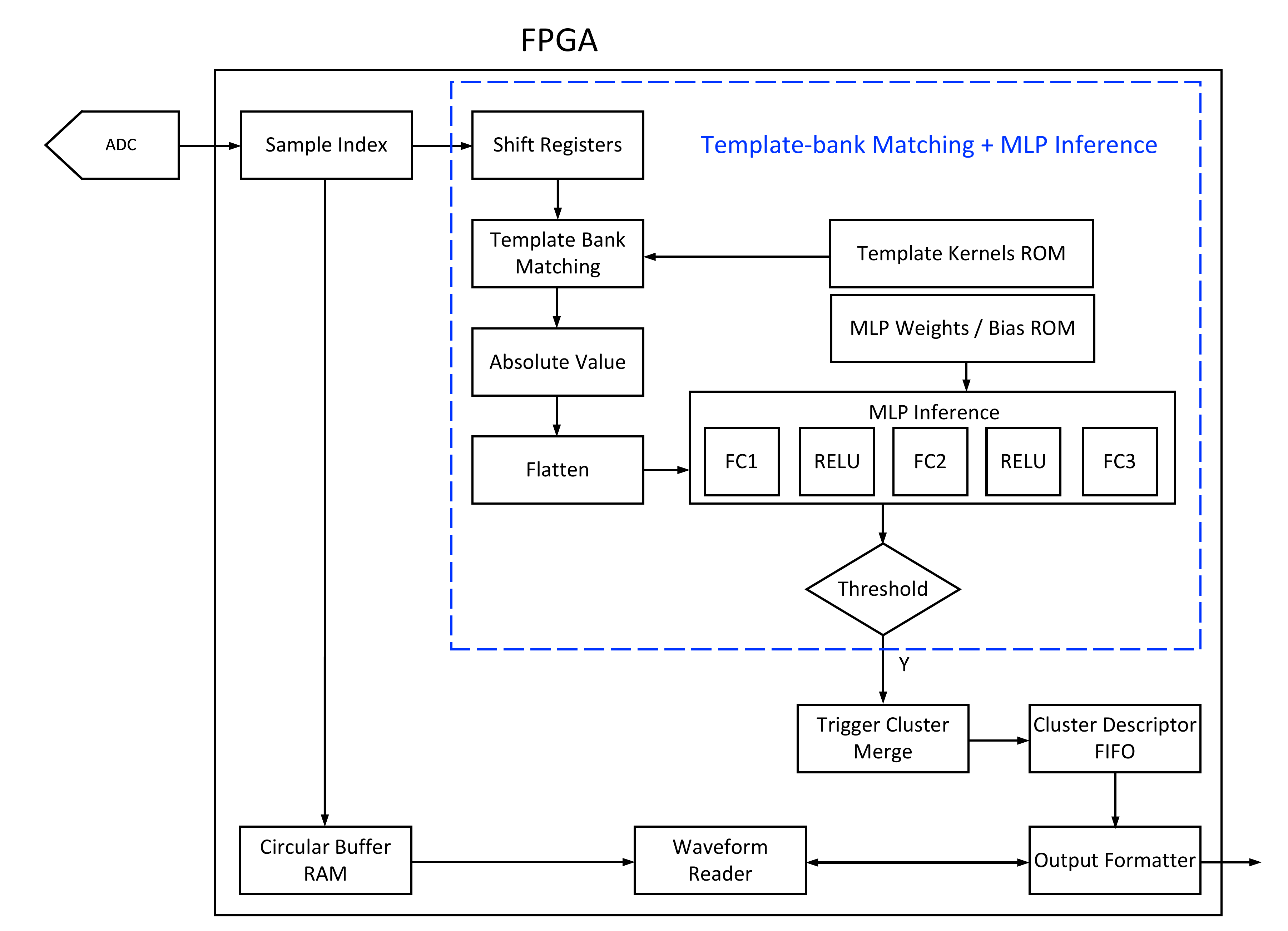}
    \caption{Overall FPGA architecture for the streaming MLP trigger. ADC: analog-to-digital converter; FC: fully connected; ROM: read-only memory.}
    \label{fig:fpga_overall_architecture_placeholder}
\end{figure}

Each input sample is assigned a sample index and sent to two paths. The waveform-cache path continuously writes the raw samples to circular random-access memory (RAM). In the trigger path, a shift register maintains the current $L=31$ sample window. The 113 \rev{OF} kernels are stored on chip and applied to this window. Their absolute responses at the current sampling time form the 113-dimensional input to the MLP trigger. The resulting output logit is compared directly with a programmable threshold to produce the trigger decision.
When the logit exceeds the threshold, the corresponding window start index is passed to the cluster-merging logic. Consecutive accepted windows are merged into one event cluster. A first-in, first-out (FIFO) queue stores the cluster boundaries and trigger metadata, and the waveform reader retrieves the corresponding raw samples from the circular RAM. The event formatter then combines the metadata and waveform samples into the record sent to the downstream data-acquisition system. Figure~\ref{fig:fpga_overall_architecture_placeholder} shows the overall FPGA data flow.

\subsection{Template-Bank Matching and MLP Evaluation}
\label{sec:fpga_parallel_implementation}

For each template $\eta$, the \rev{OF} response is computed as
\begin{equation}
\mathbf{h}_{\eta}^{\top}\mathbf{x}
=
\sum_{i=0}^{L-1} h_{\eta,i}\,x_i,
\label{eq:fpga_template_response}
\end{equation}
where $x_i$ is an input sample and $h_{\eta,i}$ is the corresponding discrete-time filter coefficient. With 113 templates and $L=31$, evaluating Eq.~\eqref{eq:fpga_template_response} requires 3503 coefficient multiplications for each new sample. The 113 signed responses are converted to their absolute values before entering the neural network.

The 113-dimensional response vector is processed by \rev{FC} layers with dimensions $113\mathbin{\to}16\mathbin{\to}8\mathbin{\to}1$, with ReLU activations after the two hidden layers. The \rev{FC} layers are evaluated with time-multiplexed multiply-accumulate (MAC) lanes.

\begin{figure*}[htp]
    \centering
    \includegraphics[width=\textwidth]{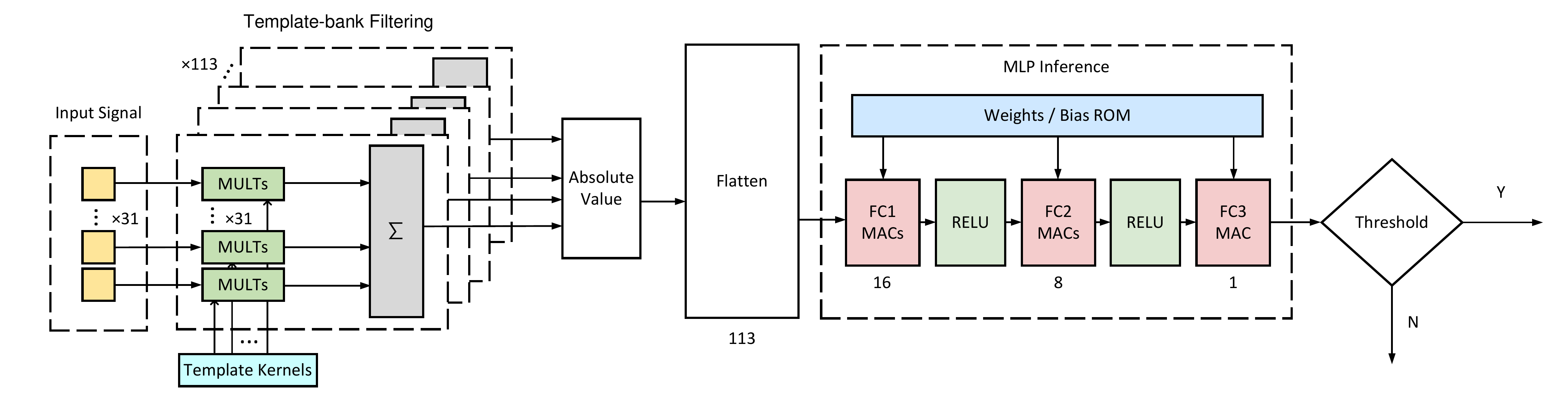}
    \caption{Internal FPGA datapath for the streaming MLP trigger. The input waveform is processed by the 113-template filter bank, converted to absolute responses, and passed through the $113\mathbin{\to}16\mathbin{\to}8\mathbin{\to}1$ MLP network and a threshold comparison. MAC: multiply-accumulate; MULT: multiplier; RELU denotes the ReLU activation.}
    \label{fig:fpga_inside}
\end{figure*}

Instantiating one multiplier for every weight would be unnecessary at the $1~\mathrm{MHz}$ input rate and would exceed the practical resource budget of the target FPGA. The design therefore uses parallel \rev{MAC} lanes that are reused across templates and neurons, consistent with the \rev{MAC} and pipelining capabilities of the 7-series digital signal processing (DSP) blocks~\cite{xilinxUG479}. Filter coefficients, network weights, and biases are stored in on-chip read-only memories (ROMs) and read in tiles~\cite{xilinxUG473}. Partial sums are accumulated for each output channel, after which the completed layer output is written to a buffer and passed to the next stage. The ReLU activations are implemented as comparisons with zero.
The internal filter and neural-network datapath is illustrated in Fig.~\ref{fig:fpga_inside}.

\subsection{On-Board Validation}
\label{sec:fpga_verification}

The on-board test waveforms are represented by signed 20-bit samples. The same quantized inputs are supplied to the bit-accurate software reference, which reproduces the fixed-point template-bank and MLP datapaths. The floating-point model provides the performance reference.

The generated bitstream is successfully programmed on the AC701 board. The trigger-core resource utilization is summarized in Table~\ref{tab:fpga_implementation_results}. The implementation meets the timing constraints at $200~\mathrm{MHz}$.

\begin{table}[htp]
    \centering
    \caption{FPGA trigger-core resource utilization. BRAM: block random-access memory; DSP: digital signal processing; LUT: look-up table.}
    \label{tab:fpga_implementation_results}
    \footnotesize\begin{tabular}{lrr}
        \hline
        Resource & Used & Utilization (\%) \\
        \hline
        LUT & 11,006 & 8.23 \\
        Register & 15,357 & 5.74 \\
        BRAM tile & 12.5 & 3.42 \\
        DSP48 & 72 & 9.73 \\
        \hline
    \end{tabular}
\end{table}

\begin{figure}[htp]
    \centering
    \includegraphics[width=0.98\columnwidth]{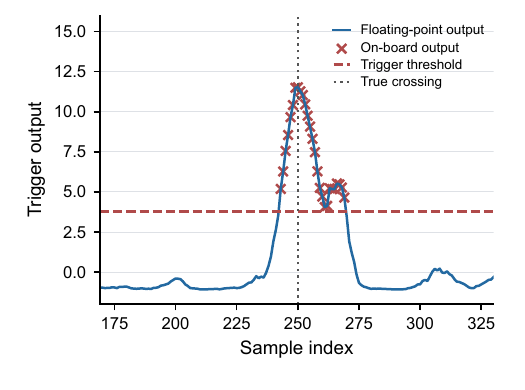}
    \caption{On-board output validation for a representative $L=31$ signal-plus-noise record. The floating-point MLP output is shown as a solid line, and the crosses mark the above-threshold FPGA outputs. The vertical dotted line indicates the true \rev{MM} crossing time, and the horizontal dashed line indicates the FPGA threshold calibrated for a stored-data fraction of $10^{-3}$.}
    \label{fig:fpga_functional_verification}
\end{figure}

Figure~\ref{fig:fpga_functional_verification} shows a representative signal-plus-noise record used in the on-board validation. All 27 above-threshold FPGA scores and their window indices agree exactly with the bit-accurate fixed-point reference. Relative to the floating-point output, the maximum absolute score difference is $4.12\times10^{-4}$.

\begin{figure}[htp]
    \centering
    \includegraphics[width=0.98\columnwidth]{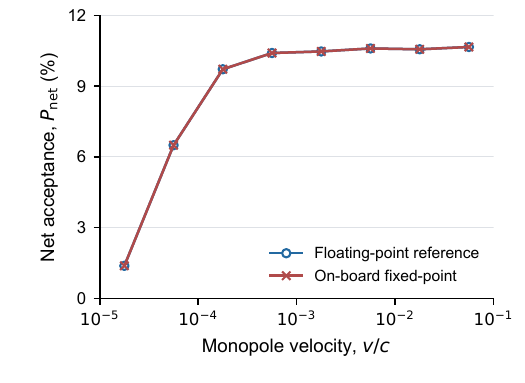}
    \caption{Velocity-resolved net acceptance $P_{\mathrm{net}}$ comparison for the $L=31$ MLP trigger at a stored-data fraction of $10^{-3}$. The floating-point reference is compared with the selected FPGA fixed-point configuration, using $3.2\times10^6$ signal-plus-noise records.}
    \label{fig:fpga_acceptance}
\end{figure}

The FPGA trigger was evaluated using the same $\mathcal{D}_{-}$ and $\mathcal{D}_{+}$ as in Section~\ref{sec:stream_evaluation}, at a stored-data fraction of $10^{-3}$. On-board tests confirmed continuous processing at $1~\mathrm{MHz}$ without dropped samples.
Figure~\ref{fig:fpga_acceptance} compares the velocity-resolved net acceptance $P_{\mathrm{net}}$ of the floating-point reference and the FPGA fixed-point implementation. The fixed-point implementation gives a net acceptance lower than the floating-point reference by only $3/(3.2\times10^6)$. The overall values of $P_{\mathrm{net}}$ are $8.7911\%$ for the floating-point reference and $8.7910\%$ for the FPGA fixed-point configuration. These tests verify the functional correctness and real-time operation of the FPGA trigger.

\section{Conclusions}
\label{sec:conclusion}

A compact template-bank construction method has been developed for the SCEP \rev{L-1} trigger using the absolute loss in detection probability as the coverage criterion. By adding the worst-covered waveform at each iteration, the method constructs a 113-template bank for $L=31$ samples, with empirical coverage above $99.99\%$ on an independent audit set. \rev{Compared with the conventional template-bank trigger, the compact template-bank trigger increases the mean $P_{\mathrm{net}}$ from $6.72\%$ to $7.52\%$ at the same stored-data fraction, an $11.9\%$ relative improvement.} To exploit the information beyond the largest template response, an MLP trigger combines the absolute responses of all templates. A waveform-window length scan then identifies $L=31$ as the configuration with the highest mean net acceptance. At a stored-data fraction of $10^{-3}$, the MLP improves $P_{\mathrm{net}}$ from $7.52\%$ to $8.79\%$, corresponding to a relative increase of $16.8\%$ over the compact template-bank trigger. The resulting network is implemented on a Xilinx AC701 FPGA platform for real-time processing of 20-bit waveforms sampled at $1~\mathrm{MHz}$, with trigger logic operating at $200~\mathrm{MHz}$. Evaluation using the same $\mathcal{D}_{-}$ and $\mathcal{D}_{+}$ gives a reduction in $P_{\mathrm{net}}$ of only $3/(3.2\times10^6)$ relative to the floating-point reference. Together, the acceptance comparison, on-board validation, and timing verification demonstrate the correctness and real-time operation of the FPGA-based \rev{L-1} trigger.

\input{acknowledgement}

\bibliographystyle{IEEEtran}
\bibliography{apssamp}

\end{document}

%% file: authorlist_neural_trigger.tex
\author{
Changqing Ye, Yunhan Wang, Yifan Qiao, Zhe Cao, Beige Liu, Qing Lin, and Lei Zhao%
\thanks{(Corresponding authors: Zhe Cao, Qing Lin, and Lei Zhao.)}%
\thanks{Changqing Ye and Beige Liu are with the Department of Modern Physics, University of Science and Technology of China, Hefei 230026, China.}%
\thanks{Yunhan Wang is with the Institute of Advanced Technology, University of Science and Technology of China, Hefei 230088, China.}%
\thanks{Yifan Qiao is with the School of Emerging Technology, University of Science and Technology of China, Hefei 230026, China.}%
\thanks{Zhe Cao, Qing Lin, and Lei Zhao are with the Department of Modern Physics, University of Science and Technology of China, Hefei 230026, China, and also with the Deep Space Exploration Laboratory, Hefei 230022, China (e-mail: caozhe@ustc.edu.cn; qinglin@ustc.edu.cn; zlei@ustc.edu.cn).}
}

%% file: acknowledgement.tex
\section*{Acknowledgements}

This project is supported by grants from National Science
Foundation of China (No. 12250011), and by the Frontier Scientific Research Program of Deep Space Exploration Laboratory under grant No. 2022-QYKYJH-HXYF-013.